\documentclass[12pt]{article}
\usepackage{epsfig}

\usepackage{amsmath}
\usepackage{amssymb}
\usepackage{graphicx}

\allowdisplaybreaks[4]

\begin{document}

\begin{titlepage}
\noindent
\begin{flushright}
{KYUSHU-HET 47}
\end{flushright}

\begin{center}
  \begin{Large}
    \begin{bf}
Matter profile effect in neutrino factory
    \end{bf}
  \end{Large}
\end{center}
\vspace{0.2cm}
\begin{center}
T.~Ota\\%
{\footnotesize \it
Department of Physics, Kyushu University, Fukuoka 812-8581, Japan
}
\\
  \vspace{0.5cm}
 J.~Sato\\%
{\footnotesize \it%
Research Center for Higher Education, Kyushu University,}\\
{\footnotesize \it%
Ropponmatsu, Chuo-ku, Fukuoka, 810-8560, Japan%
}
\end{center}

\begin{abstract}
We point out that the matter profile effect --- the effect of matter
density fluctuation on the baseline --- is very important to estimate
the parameters in a neutrino factory with a very long baseline.  To make
it clear, we propose the method of the Fourier series expansion of the
matter profile.  By using this method, we can take account of both the
matter profile effect and its ambiguity.  For very long baseline
experiment, such as $L=7332$km, in the analysis of the oscillation
phenomena we need to introduce a new parameter $ a_{1} $--- the Fourier
coefficient of the matter profile --- as a theoretical parameter to deal
with the matter profile effects.
\end{abstract}
\end{titlepage}

\section{Introduction}
The evidence of the neutrino oscillation was presented by the
Super-Kamiokande collaboration\cite{SK}.  Many experiments have been
planed and some of them have already been carried out to determine the
parameters of the neutrino oscillation\cite{Next}.  Recently, the
neutrino factory\cite{Geer} --- a very long (over 1000 km) baseline
experiment with high energy neutrinos from a muon storage ring ---has
been proposed as the most effective experiment to determine the unknown
parameters, $ U_{e3} $, the sign of $\Delta m^2$ which is responsible
for the atmospheric neutrino anomaly and the CP violating
phase\cite{BGW,Golden,BGRW,GH,DFLR,FHL,NuFact,BCR,KS,Yasuda}.

To analyze the oscillation phenomena with such a long baseline, we have
to treat the matter effect\cite{MSW} very carefully. Two ways have been
adopted to deal with the matter effects: (i) To add the averaged matter
density $ \bar{\rho} $ as a parameter in addition to the other
theoretical parameters like the mixing angles, and estimate it in the
same way as the other parameters\cite{Golden,FHL,BCR}.  (ii) To use the
PREM---Preliminary Reference Earth Model\cite{PREM}--- for the matter
density profile, and to assume that we know the ingredients of the earth
completely\cite{BGW,BGRW}.  We have, however, questions to these
treatments.  For the method-(i), ``Can we describe the matter effect
precisely enough by only averaged matter density ?'' The answer was
``No, we can't. Sometimes the deviation from the constant density is
important.  '' \cite{BGW}.  For the method-(ii), ``Is the PREM a
trustworthy model for the neutrino oscillation experiments ?''  The PREM
is originally based on the study of earthquake waves, hence it predicts
the density profile in the depth However, it doesn't predict the
ingredients of the matter. Therefore, there is an ambiguity in electron
number fraction.  Then, we should worry whether there are ambiguities in
electron number density. If we estimated the parameters without
considering its ambiguity, we would have significant errors in the
estimates for the parameters.  Therefore, we must introduce the way to
take into account the matter profile effect as the parameter
determined by the experiments.  We propose the method of the Fourier
series expansion of the matter profile\cite{KoikeSato}.  Using the
Fourier expansion we can parameterize the matter profile. We can express
the matter profile effect with a finite number of the parameters by
examining how many terms of the Fourier expansion contribute to the
oscillation physics within the resolution of the experiments. We can
incorporate the matter profile's ambiguity in the ambiguities of the
Fourier coefficients.

We will review the method of the Fourier series and investigate
qualitatively what circumstances make the matter profile effect
relevant in section 2.  In section 3, we will calculate quantitatively
the oscillation probabilities and the event rates with various sets of
the parameters in baseline lengths.

\section{The method of the Fourier series}

In this section we introduce the Fourier expansion method. First, we
describe its formalism and see its feature. Then, we solve the evolution
equation for neutrinos in matter perturbatively, and study the condition
where the matter profile effect is significant.

\subsection{Fourier expansion method}
  
Assuming three generations of neutrinos, we parameterize the lepton
mixing matrix
\begin{align}
U_{\alpha i} &\equiv e^{i \psi \lambda_{7}} \Gamma e^{i \phi \lambda_{5}}
e^{i \omega \lambda_{2}} U_{\mathrm{Majorana}} \nonumber \\
&=\begin{pmatrix}
     1 &    & \\
       & c_{\psi}  & s_{\psi} \\
       & -s_{\psi} & c_{\psi}
   \end{pmatrix}
   \begin{pmatrix}
     1 &   &  \\
       & 1 &  \\
       &   & e^{i \delta}
   \end{pmatrix}
   \begin{pmatrix}
     c_{\phi} &   & s_{\phi} \\
              & 1 &          \\
     -s_{\phi}&   & c_{\phi}
   \end{pmatrix}
   \begin{pmatrix}
     c_{\omega} & s_{\omega} & \\
     -s_{\omega}& c_{\omega} & \\
                &            & 1
   \end{pmatrix} U_{\mathrm{Majorana}} \nonumber \\
 &=\begin{pmatrix}
     c_{\phi} c_{\omega}
      & c_{\phi} s_{\omega}
       & s_{\phi} \\
     - c_{\psi} s_{\omega} - s_{\psi} s_{\phi} c_{\omega}
       e^{i \delta}
      & c_{\psi} c_{\omega} - s_{\psi} s_{\phi} s_{\omega}
       e^{i \delta}
       & s_{\psi} c_{\phi} e^{i \delta} \\
     s_{\psi} s_{\omega} - c_{\psi} s_{\phi} c_{\omega}
       e^{i \delta}
      & -s_{\psi} c_{\omega} - c_{\psi} s_{\phi} s_{\omega}
       e^{i \delta}
       & c_{\psi} c_{\phi} e^{i \delta}
    \end{pmatrix}
    U_{\mathrm{Majorana}}.
\end{align}
which relates the flavor eigenstates $ | \nu_{\alpha}
\rangle (\alpha = e, \mu, \tau) $ with the mass eigenstates in vacuum $
| \nu_{i} \rangle (i = 1, 2, 3) $ as
\begin{eqnarray}
| \nu_{\alpha} \rangle = U_{\alpha i} | \nu_{i} \rangle.
\end{eqnarray}
$ U_{\mathrm{Majorana}} $ is the part of the Majorana phases. It doesn't
contribute to the neutrino oscillation phenomena and hence is omitted
hereafter.

The evolution in matter of the flavor eigenstates of
neutrinos\footnote{For antineutrino $a(x)$ and $\delta$ should be
replaced by $-a(x)$ and $-\delta$.} with its
energy $E$ is given by
\begin{eqnarray}
i \frac{d}{dx} | \nu_{\beta} (x) \rangle
 &=& H(x)_{\beta \alpha} | \nu_{\alpha} (x) \rangle 
\label{Schrodinger eq}\\
H(x)_{\beta \alpha} &\equiv&
\frac{1}{2 E}
     \left\{
       U_{\beta i}
        \begin{pmatrix}
            0 & & \\
              & \Delta m_{21}^{2} & \\
              &                   & \Delta m_{31}^{2}
        \end{pmatrix}
       U^{\dagger}_{i \alpha}
                +
        \begin{pmatrix}
         a(x) &   & \\
              & 0 & \\
              &   & 0
        \end{pmatrix}_{\beta \alpha}
       \right\},
\nonumber 
\end{eqnarray}
where
\begin{gather}
\Delta m_{ij}^{2}\equiv m_{i}^{2} - m_{j}^{2},
\quad m_{i}:\text{mass eigenvalue} \nonumber \\
a(x) \equiv 2 \sqrt{2} G_{F} n_{e} (x) E=7.56
 \times 10^{-5} \left( \frac{\rho(x)}{\mathrm{g/cm^{3}}} \right)
\left( \frac{Y_e}{0.5} \right)
 \left( \frac{E}{\mathrm{GeV}} \right) [\mathrm{eV^{2}}]  \\
\begin{array}{lcl}
G_F&:&\text{Fermi constant}\\
 n_{e} (x)&:&\text{electron number density}  \\
 \rho (x)&:&\text{matter density}\\
Y_e&:&\text{electron fraction}
\end{array}
\nonumber
\end{gather}
To parameterize the matter effect, $a(x)$, we expand $a(x)$ into Fourier
series as
\begin{gather}
a (x) = \sum_{n = - \infty}^{\infty} a_{n} e^{-i p_{n} x},
\qquad
p_{n} \equiv \frac{2 \pi}{L} n .
\end{gather} 
Note that $ a_{-n} = a_{n}^{*} $  due to the reality of $ a(x) $.
Also within the PREM,
\begin{eqnarray}
a_{-n}=a_{n}
\label{an Condition}
\end{eqnarray}
since $a(x)=a(L-x)$.

If this expansion can  be approximated with a finite number ($\equiv N$)
of terms,
\begin{eqnarray}
 a(x)=\sum_{n=-N}^{n=N} a_{n} e^{-i p_{n} x},
\label{Fourier Profile}
\end{eqnarray}
then it means that the matter profile effect can be parametrized.
Thus, by introducing new parameters $a_n (n=-N,\cdots ,N)$
(or $n=0,\cdots ,N$, if eq.(\ref{an Condition}) holds), we can
investigate the oscillation physics without help of earth models.

\subsection{Perturbative analysis of matter profile effect}

We solve the evolution equation (\ref{Schrodinger eq}) to see the
qualitative feature of the matter profile effect. First, we divide the
Hamiltonian $H(x)$ into three pieces for the later calculation:
\begin{align}
H_{00} &\equiv \frac{1}{2 E}
        e^{i \psi \lambda_{7}} \Gamma
          \begin{pmatrix}
           \Delta m_{31}^{2} s_{\phi}^{2}
            + \bar{a}
             + \Delta m_{21}^{2} c_{\phi}^{2} s_{\omega}^{2}
            & 0
             & \Delta m_{31}^{2} c_{\phi} s_{\phi}
               -\Delta m_{21}^{2} c_{\phi} s_{\phi} s_{\omega}^{2}
                \\
           0 & \Delta m_{21} c_{\omega}^{2} & 0 \\
           \Delta m_{31}^{2} c_{\phi} s_{\phi}
            -\Delta m_{21}^{2} c_{\phi} s_{\phi} s_{\omega}^{2}
            & 0
             & \Delta m_{31}^{2} c_{\phi}^{2}
               + \Delta m_{21} s_{\phi}^{2} s_{\omega}^{2}
          \end{pmatrix}
         \Gamma^{\dagger} e^{-i \psi \lambda_{7}} ,\\
H_{01} &\equiv \frac{\Delta m_{21}^{2}}{2 E}
         c_{\omega} s_{\omega}
          e^{i \psi \lambda_{7}} \Gamma
           \begin{pmatrix}
            0        & c_{\phi}  & 0 \\
            c_{\phi} &    0      & -s_{\phi} \\
            0        & -s_{\phi} & 0
           \end{pmatrix}
          \Gamma^{\dagger} e^{-i \psi \lambda_{7}} , \\
H_{1}(x) &\equiv \frac{1}{2 E}
        \begin{pmatrix}
         \delta a(x) &   & \\
                     & 0 & \\
                     &   & 0
        \end{pmatrix},
\end{align}
where $\bar{a}\equiv a_0$ and $\delta a(x) \equiv a(x)-\bar{a}$.
The oscillation phenomena arise from $H_{00}$, hence its effect must
be fully taken into account. On the other hand $H_{01}$ and $H_1(x)$ can
be treated as the perturbation in almost all cases in neutrino factories,
since
\begin{eqnarray}
 \left| \frac{\Delta m^2_{21}L}{2E} \right| \ll 1 \quad \text{ and }
\quad \left| \frac{\delta a(x)L}{2E} \right| \ll 1
\label{perturbativeCondition}
\end{eqnarray}
can be almost always expected. Note that we have to include $\Delta
m^2_{21}$ terms in $(H_{00})_{ij} (i,j=1,3)$ to deal with the resonance
effect.\footnote{On the other hand, we include $\Delta m^2_{21}$ term in
the (22) element just for convenience. This inclusion is not essential.}

Next, we rewrite $H_{00}$ as
\begin{align}
H_{00} &= \frac{1}{2 E}
       \tilde{U}_{0}
        \begin{pmatrix}
         \lambda_{-} & & \\
                     & \Delta m_{21}^{2} c_{\omega}^{2} & \\
                     & & \lambda_{+}
        \end{pmatrix}
       \tilde{U}_{0}^{\dagger} 
\end{align}
where $ \tilde{U}_{0} $ and $ \lambda_{\pm} $ are
the mixing matrix and mass square eigenvalues in matter
up to the zeroth order of the ``perturbation''s $H_{01}$ and $H_1(x)$
and defined by
\begin{eqnarray}
\tilde{U}_{0} &\equiv&
  e^{i \psi \lambda_{7}} \Gamma
   e^{i \tilde{\phi} \lambda_{5}}
  =\begin{pmatrix}
     c_{\tilde{\phi}}
      & 0
       & s_{\tilde{\phi}} \\
     - s_{\psi} s_{\tilde{\phi}} e^{i \delta}
      & c_{\psi}
       & s_{\psi} c_{\tilde{\phi}} e^{i \delta} \\
     - c_{\psi} s_{\tilde{\phi}} e^{i \delta}
      & s_{\psi}
       & c_{\psi} c_{\tilde{\phi}} e^{i \delta}
    \end{pmatrix} \\
\tan 2 \tilde{\phi} &=& \frac{s_{2 \phi} ( \Delta m_{31}^{2} - \Delta
m_{21}^{2} s_{\omega}^{2})}{c_{2 \phi} ( \Delta m_{31}^{2} - \Delta
m_{21}^{2} s_{\omega}^{2}) - \bar{a}},
\\
\lambda_{\pm} &=& \frac{1}{2}\{\alpha \pm \beta\},
\\
 &\alpha &\equiv
     \Delta m_{31}^{2} + \Delta m_{21}^{2} s_{\omega}^{2} + \bar{a},
\nonumber \\
&\beta&\equiv
     \sqrt{ \left\{
            ( \Delta m_{31}^{2} - \Delta m_{21}^{2} s_{\omega}^
                {2} ) c_{2 \phi} - \bar{a}
            \right\}^{2}
            +
            ( \Delta m_{31}^{2} - \Delta m_{21}^{2} s_{\omega}^
             {2} )^{2} s_{2 \phi}^{2}
           } \nonumber .
\end{eqnarray}


Then we solve the evolution equation (\ref{Schrodinger eq})
perturbatively as
\begin{align}
| \nu_{\beta} (L) \rangle
 &= \mathrm{T}
      \left[
      \exp \left( -i \int_{0}^{L}  H(x) dx \right)
      \right]_{\beta \alpha}
     | \nu_{\alpha} (0) \rangle \nonumber \\
 &= \left( e^{-i H_{00} L} +
     e^{-i H_{00} L} (-i) \int_{0}^{L} {H_{01}}_{I} dx + 
     e^{-i H_{00} L} (-i) \int_{0}^{L} {H_{1}(x)}_{I} dx + \dotsb
    \right)_{\beta \alpha} | \nu_{\alpha} (0) \rangle
  \nonumber \\
 & \equiv ( S_{00} + S_{01} + S_{1} + \dotsb)_{\beta \alpha} 
| \nu_{\alpha} (0) 
   \rangle,
\end{align}
where $ S_{00},S_{01} $ and $ S_{1} $ are defined by the first, second
and third terms in the second last line respectively. $S_{01}$ and $S_1$
are the perturbative contributions at the first order from $H_{01}$ and
$H_1$ respectively.  $ S_{00} $, $ S_{01} $ and $ S_{1} $ are calculated
to be
\begin{align}
S_{00}(L)_{\beta \alpha}
  &\equiv e^{-i H_{00} L} \nonumber \\
 &= \tilde{U}_{0}
      \begin{pmatrix}
       e^{-i \frac{\lambda_{-}}{2 E} L} & & \\
        & e^{-i \frac{\Delta m_{21}^{2} c_{\omega}^{2}}{2
         E} L} &
          \\
        & & e^{-i \frac{\lambda_{+}}{2 E} L}
      \end{pmatrix}
      \tilde{U}_{0}^{\dagger} \nonumber \\
 &= \left(
     \begin{array}{cc}
      s_{\tilde{\phi}}^{2} e^{-i \frac{\lambda_{+}}{2E} L}
       + c_{\tilde{\phi}}^{2} e^{-i \frac{\lambda_{-}}{2E} L}
     & \frac{1}{2} e^{-i \delta} s_{\psi} s_{2 \tilde{\phi}}
        ( e^{-i \frac{\lambda_{+}}{2E} L}
          - e^{-i \frac{\lambda_{-}}{2E} L} )
     \\
      \frac{1}{2} e^{i \delta} s_{\psi} s_{2 \tilde{\phi}}
        ( e^{-i \frac{\lambda_{+}}{2E} L}
          - e^{-i \frac{\lambda_{-}}{2E} L} )
     & s_{\psi}^{2} (c_{\tilde{\phi}}^{2} e^{-i \frac{\lambda_
     {+}}
      {2E} L} +s_{\tilde{\phi}}^{2} e^{-i \frac{\lambda_{-}}
      {2E} L} )+
       c_{\psi}^{2}
        e^{-i \frac{\Delta m_{21} c_{\omega}^{2}}{2E} L}
     \\
     \frac{1}{2} e^{i \delta} c_{\psi} s_{2 \tilde{\phi}}
        ( e^{-i \frac{\lambda_{+}}{2E} L}
          - e^{-i \frac{\lambda_{-}}{2E} L} )
     & \frac{1}{2} s_{2 \psi}
      (c_{\tilde{\phi}}^{2} e^{-i \frac{\lambda_{+}}{2E} L}
       +s_{\tilde{\phi}}^{2} e^{-i \frac{\lambda_{-}}{2E} L}
       -e^{-i \frac{\Delta m_{21} c_{\omega}^{2}}{2E} L})
     \end{array}
     \right. \nonumber \\
 & \left.
    \begin{array}{cc}
     \qquad \qquad \qquad \qquad \qquad \qquad
      & \frac{1}{2} e^{-i \delta} c_{\psi} s_{2 \tilde{\phi}}
         ( e^{-i \frac{\lambda_{+}}{2E} L}
           - e^{-i \frac{\lambda_{-}}{2E} L} )
      \\
     \qquad \qquad \qquad \qquad \qquad \qquad
      & \frac{1}{2} s_{2 \psi}
      (c_{\tilde{\phi}}^{2} e^{-i \frac{\lambda_{+}}{2E} L}
       +s_{\tilde{\phi}}^{2} e^{-i \frac{\lambda_{-}}{2E} L}
       -e^{-i \frac{\Delta m_{21} c_{\omega}^{2}}{2E} L})
      \\
     \qquad \qquad \qquad \qquad \qquad \qquad
      & c_{\psi}^{2} (c_{\tilde{\phi}}^{2} e^{-i \frac{\lambda_
      {+}}
      {2E} L} + s_{\tilde{\phi}}^{2} e^{-i \frac{\lambda_{-}}{2E}
        L} ) + s_{\psi}^{2} e^{-i \frac{\Delta m_{21} c_{\omega}^
        {2}}{2E} L}
    \end{array}
    \right),
\label{S00} \\
S_{01}(L)_{\beta \alpha}
 &\equiv e^{-i H_{00} L}
         (-i) \int_{0}^{L} dx {H_{01}}_{I} \nonumber \\
 &= \tilde{U}_{0}
     \begin{pmatrix}
      0            & \mathcal{A} & 0      \\
      \mathcal{A}  & 0           & \mathcal{B} \\
      0            & \mathcal{B} & 0
     \end{pmatrix}
     \tilde{U}_{0}^{\dagger} \nonumber \\
 &=
  \begin{pmatrix}
   0 & c_{\psi}(c_{\tilde{\phi}} \mathcal{A} + s_{\tilde{\phi}} \mathcal{B})
     & -s_{\psi}(c_{\tilde{\phi}} \mathcal{A} + s_{\tilde{\phi}} \mathcal{B}) \\
   c_{\psi}(c_{\tilde{\phi}} \mathcal{A} + s_{\tilde{\phi}} \mathcal{B})
    & - s_{2 \psi}(s_{\tilde{\phi}} \mathcal{A} - c_{\tilde{\phi}}
      \mathcal{B})
       c_{\delta}
    & (e^{i \delta}s_{\psi}^{2} - e^{-i \delta}c_{\psi}^{2})
       (s_{\tilde{\phi}} \mathcal{A} - c_{\tilde{\phi}} \mathcal{B}) \\
   -s_{\psi}(c_{\tilde{\phi}} \mathcal{A} + s_{\tilde{\phi}} \mathcal{B})
    & -(e^{i \delta}c_{\psi}^{2} - e^{-i \delta}s_{\psi}^{2})
       (s_{\tilde{\phi}} \mathcal{A} - c_{\tilde{\phi}} \mathcal{B})
    & s_{2 \psi}(s_{\tilde{\phi}} \mathcal{A} - c_{\tilde{\phi}}
     \mathcal{B})
       c_{\delta}
  \end{pmatrix}
\label{S01}
\end{align}
\begin{gather}
\mathcal{A} \equiv
  \frac{ \Delta m_{21}^{2} }
   {2 ( \Delta m_{21}^{2} c_{\omega}^{2} - \lambda_{-} )}
  c_{\phi - \tilde{\phi}} s_{2 \omega}
  \left(
   e^{-i \frac{\Delta m_{21}^{2} c_{\omega}^{2}}{2 E} L}
   -
   e^{-i \frac{\lambda_{-}}{2 E} L}
  \right) \\
\mathcal{B} \equiv
  \frac{ \Delta m_{21}^{2} }
   {2 ( \lambda_{+} - \Delta m_{21}^{2} c_{\omega}^{2} )}
  s_{\phi - \tilde{\phi}} s_{2 \omega}
  \left(
   e^{-i \frac{\Delta m_{21}^{2} c_{\omega}^{2}}{2 E} L}
   -
   e^{-i \frac{\lambda_{+}}{2 E} L}
  \right).
\end{gather}
\begin{align}
&S_{1} (L)_{\beta \alpha} \nonumber\\
\equiv& e^{-i H_{00} L}
         (-i) \int_{0}^{L} dx {H_{1}(x)}_{I} \nonumber \\
 = &\sum_{n}
    \tilde{U}_{0}
     \begin{pmatrix}
      0               & 0 & \mathcal{C}_{n} \\
      0               & 0 & 0     \\
      \mathcal{D}_{n} & 0 & 0
     \end{pmatrix}
    \tilde{U}_{0}^{\dagger} \nonumber \\
 = &\sum_{n}
    \begin{pmatrix}
     (\mathcal{C}_{n} + \mathcal{D}_{n}) s_{\tilde{\phi}} c_{\tilde{\phi}}
      & e^{-i \delta} s_{\psi} (- \mathcal{D}_{n} s_{\tilde{\phi}}^{2}
         +\mathcal{C}_{n} c_{\tilde{\phi}}^{2})
      & e^{-i \delta} c_{\psi} (- \mathcal{D}_{n} s_{\tilde{\phi}}^{2}
         +\mathcal{C}_{n} c_{\tilde{\phi}}^{2}) \\
      e^{i \delta} s_{\psi} ( \mathcal{D}_{n} c_{\tilde{\phi}}^{2}
         +\mathcal{C}_{n} s_{\tilde{\phi}}^{2})
      & - (\mathcal{C}_{n} + \mathcal{D}_{n}) s_{\psi}^{2} s_{\tilde{\phi}}
        c_{\tilde{\phi}}
      &  - (\mathcal{C}_{n} + \mathcal{D}_{n}) s_{\psi} c_{\psi}
       s_{\tilde{\phi}} c_{\tilde{\phi}} \\
      e^{i \delta} c_{\psi} ( \mathcal{D}_{n} c_{\tilde{\phi}}^{2}
         +\mathcal{C}_{n} s_{\tilde{\phi}}^{2})
      & - (\mathcal{C}_{n} + \mathcal{D}_{n}) s_{\psi} c_{\psi}
       s_{\tilde{\phi}} c_{\tilde{\phi}}
      & - (\mathcal{C}_{n} + \mathcal{D}_{n}) c_{\psi}^{2} s_{\tilde{\phi}}
        c_{\tilde{\phi}}
    \end{pmatrix} \nonumber \\
 = &\sum_{n} a_{n}
     \frac{s_{2 \tilde{\phi}}
      ( e^{-i \frac{ \lambda_{+} }{2E} L}
        - e^{-i \frac{ \lambda_{-} }{2E} L} )}
      { 2 \left\{
          ( \lambda_{+}
           - \lambda_{-} )^{2} - ( 2E p_{n} )^{2}
          \right\} }
\label{S1}       \\
  \times&
    \begin{pmatrix}
     ( \lambda_{+} - \lambda_{-} )
      s_{2 \tilde{\phi}}
     & e^{-i \delta} s_{\psi}
       \left\{
        ( \lambda_{+} - \lambda_{-} ) c_{2 \tilde{\phi}}
        - 2E p_{n}
       \right\}
     & e^{-i \delta} c_{\psi}
       \left\{
        ( \lambda_{+} - \lambda_{-} ) c_{2 \tilde{\phi}}
        - 2E p_{n}
       \right\} \\
     e^{i \delta} s_{\psi}
       \left\{
        ( \lambda_{+} - \lambda_{-} ) c_{2 \tilde{\phi}}
        + 2E p_{n}
       \right\}
     & - ( \lambda_{+} - \lambda_{-} )
        s_{\psi}^{2} s_{2 \tilde{\phi}}
     & \frac{1}{2}
       ( \lambda_{+} - \lambda_{-} )
       s_{2 \psi} s_{2 \tilde{\phi}} \\
     e^{i \delta} c_{\psi}
       \left\{
        ( \lambda_{+} - \lambda_{-}) c_{2 \tilde{\phi}}
        + 2E p_{n}
       \right\}
     & \frac{1}{2}
       ( \lambda_{+} - \lambda_{-} )
       s_{2 \psi} s_{2 \tilde{\phi}}
     & - ( \lambda_{+} - \lambda_{-} )
        c_{\psi}^{2} s_{2 \tilde{\phi}}
    \end{pmatrix},
\nonumber
\end{align}
\begin{gather}
\mathcal{C}_{n} \equiv
  \frac{a_{n}}{2 ( \lambda_{+} - \lambda_{-} + 2 E p_{n} )}
   s_{2 \tilde{\phi}}
    \left(
     e^{-i \frac{\lambda_{+}}{2 E} L}
      -
     e^{-i \frac{\lambda_{-}}{2 E} L}
    \right) =\mathcal{D}_{-n} \\
\mathcal{D}_{n} \equiv
  \frac{a_{n}}{2 ( \lambda_{+} - \lambda_{-} - 2 E p_{n} )}
   s_{2 \tilde{\phi}}
    \left(
     e^{-i \frac{\lambda_{+}}{2 E} L}
      -
     e^{-i \frac{\lambda_{-}}{2 E} L}
    \right) =\mathcal{C}_{-n}.
\end{gather}

With eq.(\ref{S00}), (\ref{S01}) and (\ref{S1}), the oscillation
probability, for example from $ \nu_{e} $ to $ \nu_{\mu} $, up to the lowest
order of the perturbations, $H_{01}$ and $H_{1}(x)$, is calculated as
\begin{align}
P_{\nu_{e} \rightarrow \nu_{\mu}}
 &= | S(L)_{\mu e} |^{2} \nonumber \\
 &\simeq | (S_{00})_{\mu e} |^{2}
   + 2 Re [ (S_{00})_{\mu e} (S_{01})_{\mu e}^{*} ]
   + 2 Re [ (S_{00})_{\mu e} (S_{1})_{\mu e}^{*} ] \nonumber \\
 &= s_{\psi}^{2} s_{2 \tilde{\phi}}^{2}
   \sin^{2} \frac{\lambda_{+}-\lambda_{-}}{4E} L \nonumber \\
 & \qquad +
  \frac{1}{2} c_{\delta} s_{2 \psi} s_{2 \omega} s_{2 \tilde
   {\phi}} \nonumber \\
 & \qquad \qquad \times \Biggl[
    \left( c_{\tilde{\phi}} c_{\phi - \tilde{\phi}}
    \frac{\Delta m_{21}^{2}}{\Delta m_{21}^{2} c_{\omega}^
     {2}-\lambda_{-}}
      +
      s_{\tilde{\phi}} s_{\phi - \tilde{\phi}}
    \frac{\Delta m_{21}^{2}}{\lambda_{+}-\Delta m_{21}^{2}
    c_{\omega}^{2}}
    \right)
    \sin^{2} \frac{\Delta m_{21}^{2} c_{\omega}^
     {2}-\lambda_{-}}{4E} L \nonumber \\
 & \qquad \qquad \quad -
  \left( c_{\tilde{\phi}} c_{\phi - \tilde{\phi}}
    \frac{\Delta m_{21}^{2}}{\Delta m_{21}^{2} c_{\omega}^
     {2}-\lambda_{-}}
      +
      s_{\tilde{\phi}} s_{\phi - \tilde{\phi}}
    \frac{\Delta m_{21}^{2}}{\lambda_{+}-\Delta m_{21}^{2}
    c_{\omega}^{2}}
    \right)
    \sin^{2} \frac{\lambda_{+}-\Delta m_{21}^{2} c_{\omega}^
     {2}}{4E} L \nonumber \\
 & \qquad \qquad \quad +
  \left( c_{\tilde{\phi}} c_{\phi - \tilde{\phi}}
    \frac{\Delta m_{21}^{2}}{\Delta m_{21}^{2} c_{\omega}^
     {2}-\lambda_{-}}
      -
      s_{\tilde{\phi}} s_{\phi - \tilde{\phi}}
    \frac{\Delta m_{21}^{2}}{\lambda_{+}-\Delta m_{21}^{2}
    c_{\omega}^{2}}
    \right)
    \sin^{2} \frac{\lambda_{+}-\lambda_{-}}{4E} L
    \Biggr] \nonumber \\
 & \qquad +\frac{1}{4}
  s_{\delta} s_{2 \psi} s_{2 \omega} s_{2 \tilde{\phi}}
  \left( c_{\tilde{\phi}} c_{\phi - \tilde{\phi}}
    \frac{\Delta m_{21}^{2}}{\Delta m_{21}^{2} c_{\omega}^
     {2}-\lambda_{-}}
      +
      s_{\tilde{\phi}} s_{\phi - \tilde{\phi}}
    \frac{\Delta m_{21}^{2}}{\lambda_{+}-\Delta m_{21}^{2}
    c_{\omega}^{2}}
    \right) \nonumber \\
 & \qquad \qquad \times
  \left(
   \sin \frac{\lambda_{+}-\Delta m_{21}^{2} c_{\omega}^
     {2}}{2E} L
   + \sin \frac{\Delta m_{21}^{2} c_{\omega}^
     {2}-\lambda_{-}}{2E} L
   - \sin \frac{\lambda_{+}-\lambda_{-}}{2E} L
  \right) \nonumber \\
 & \qquad +
  2 s_{\psi}^{2} s_{2 \tilde{\phi}}^{2}
  \sum_{n} a_{n}^{*}
   \frac{(\lambda_{+}-\lambda_{-}) c_{2 \tilde{\phi}}+2E p_
   {n}}{(\lambda_{+}-\lambda_{-})^{2} - (2E p_{n})^{2}}
  \sin^{2} \frac{\lambda_{+} - \lambda_{-}}{4E} L.
\label{1st order probability}
\end{align}
From this equation (\ref{1st order probability}) we find the following;
(i) The matter
profile effect is relevant when some of the Fourier coefficients $ a_{n}
$'s are as large as $ \Delta m_{21}^{2} $ or the resonance
condition\cite{Parametric},
\begin{eqnarray}
\lambda_{+} - \lambda_{-} = 2 E p_{n} ,
\label{Resonance Condition}
\end{eqnarray}
is satisfied. (ii) The matter profile effect decreases in proportion to
$ 1 / n $ when the resonance condition is not satisfied. Therefore the
higher Fourier modes are expected to be irrelevant. In other words, we
don't need the detail of the matter profile.  Thus we expect that we can
truncate the Fourier expansion of $a(x)$ at the finite number $N$.  We
can estimate the matter profile effect in neutrino factories.

\section{Numerical analysis}

In this section we examine how many terms in the Fourier expansion are
necessary. To see this we compare the oscillation probability and the
event rate calculated using the PREM with those calculated using the
truncated Fourier series (\ref{Fourier Profile}). We first calculate the
transition probabilities with several matter profiles, the PREM matter
profile and the Fourier series matter profiles that are truncated at
various $N$'s. Next using these probabilities we derive the event rates
in the appearance channel, $\nu_e\rightarrow\nu_\mu$ as the ``observed''
numbers in an experiment. To derive them we use the following equation
with a given number of muons in decay $ N_{\mu} $, muon energy $ E_{\mu}
$, detector size $ N_{kt} $ and detection efficiency $ \epsilon $.
\begin{align}
N_{\nu_e \rightarrow \nu_\mu}
 &= \int_{E_{th}}^{E_{\mu}} \frac{dE}{E_{\mu}}
    \times 
     \overbrace{
      \frac{N_{\mu}}{\pi L^{2}} \gamma^{2} 
       12 \left(  \frac{E}{E_{\mu}} \right)^{2}
       \left( 1 - \frac{E}{E_{\mu}} \right)}^{flux}
    \times 
     \overbrace{P_{\nu_{e} \rightarrow \nu_{\mu}}(E)}^{probability}
    \times 
      \overbrace{\sigma_{\nu}(E) N_{kt} N_{A} \epsilon}^{detector} 
 \nonumber
\end{align}
\begin{gather}
\sigma_{\nu}=0.67 \times 10^{-42} 
             \times \left( \frac{E}{\mathrm{GeV}} \right)[\mathrm{m^{2}}]
            :\text{CC cross section} \nonumber \\
N_{A}:\text{Avogadro's number} \nonumber \\
\gamma \equiv \frac{E_{\mu}}{m_{\mu}}, \quad m_{\mu}:\text{muon mass} 
\nonumber \\
E_{th}:\text{threshold energy of the detection}. \nonumber
\end{gather}
Then we compare these results. Through these comparisons we find how
many Fourier coefficients we have to introduce to the theoretical
parameters.

In the following we assume that $N_\mu \times N_{\rm kt} \times
\epsilon$ equals to $=5\times 10^{22}$ kt.\footnote{We also assume that
we can reconstruct neutrino energy.}

\subsection{$L=3000$km}
\begin{figure}[tbp]
\unitlength 1cm
\begin{picture}(16,10)(0,0)
\put(0,9.5){\Large$\rho$ [g/cm$^3$]}
\put(6,-0.6){(a) Matter profile}
\put(15.2,0){[km]}
\includegraphics[width=15cm]{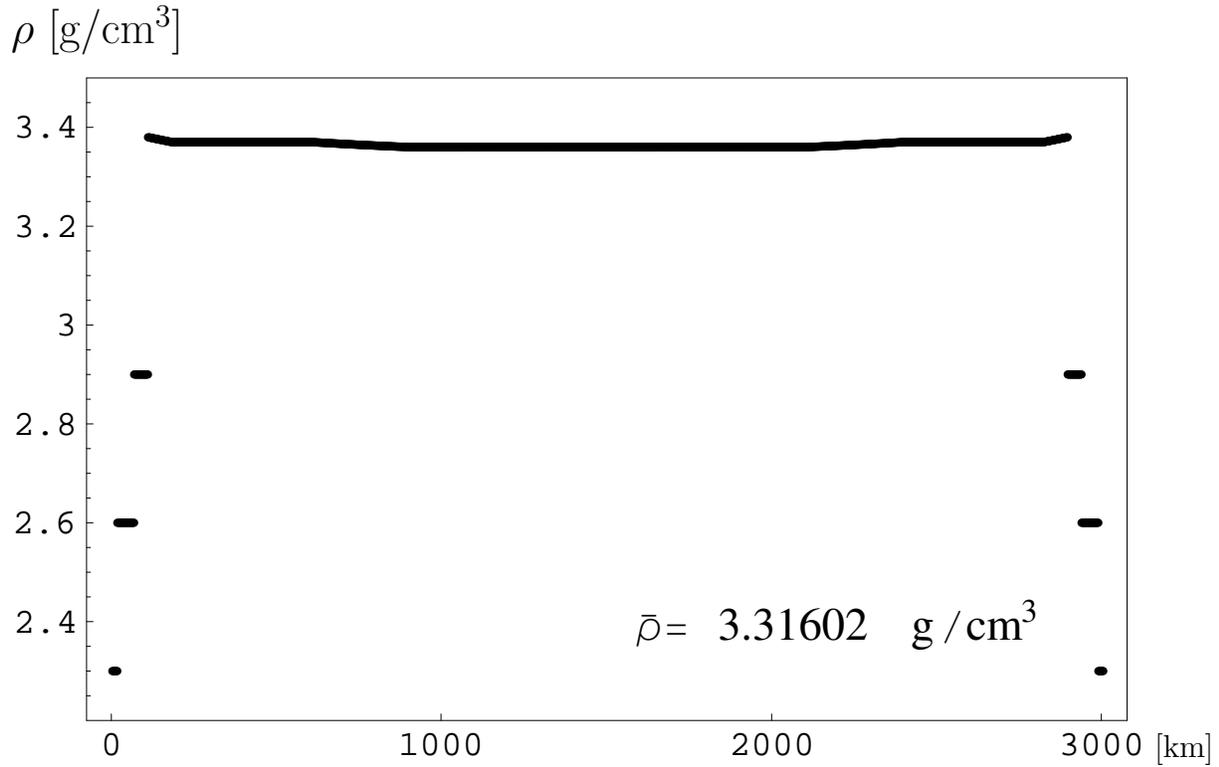}
\end{picture}
\begin{picture}(16,10.5)(0,0)
\put(0,9.5){\Large$\rho_n$ [g/cm$^3$]}
\put(6,-0.6){(b) Fourier Coeeficients}
\put(15.2,0){(mode)}
\includegraphics[width=15cm]{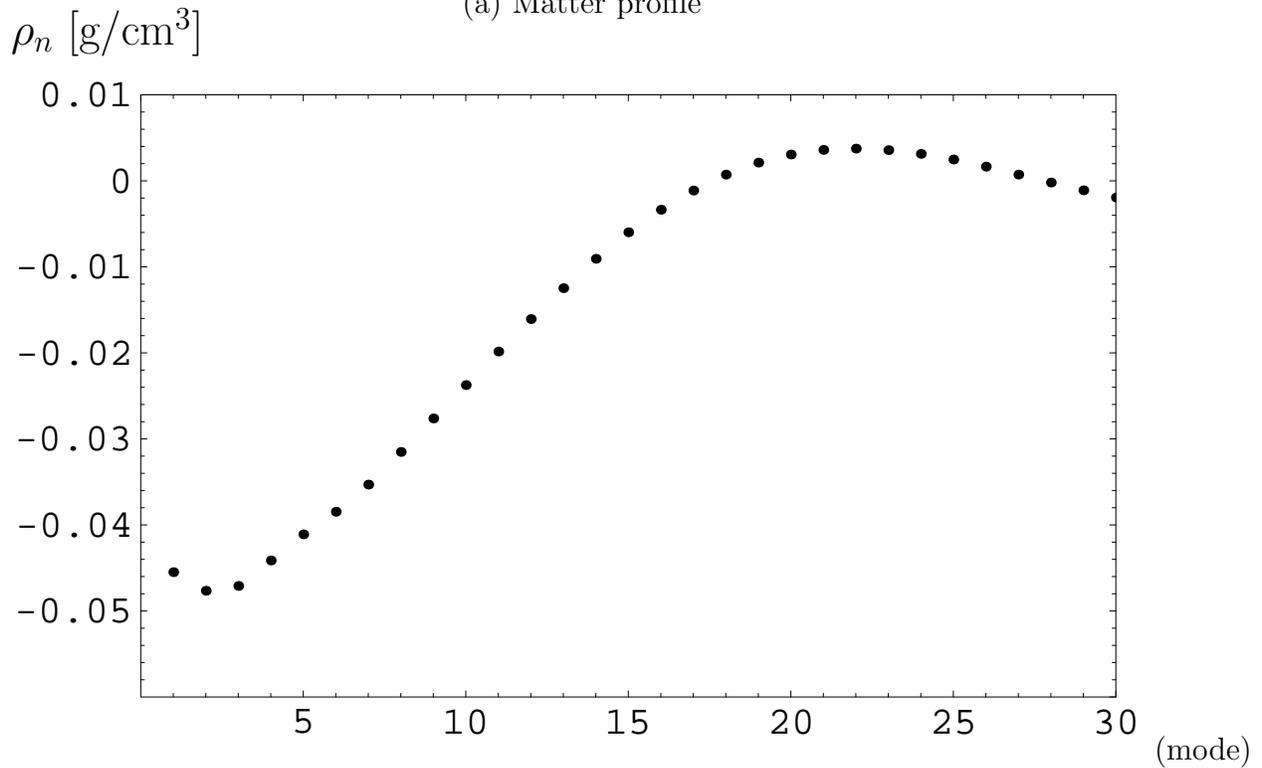} 
\end{picture}
\\
\vspace*{0.5cm}
\caption{(a) Matter profile on the baseline and (b) its Fourier
coefficients for $L=3000$km.}
\label{3000 matter profile}
\end{figure}

Neutrino factories of this length are considered to be most efficient, so
we consider this case first.

\begin{figure}[pht]
\unitlength 1cm
\begin{picture}(16,10)(0,0)
\put(0,9.5){\LARGE$\displaystyle P_{\nu_e\rightarrow\nu_\mu}$}
\put(6.8,-0.8){\LARGE$\displaystyle E_\nu$ [GeV]}
\includegraphics[width=15cm]{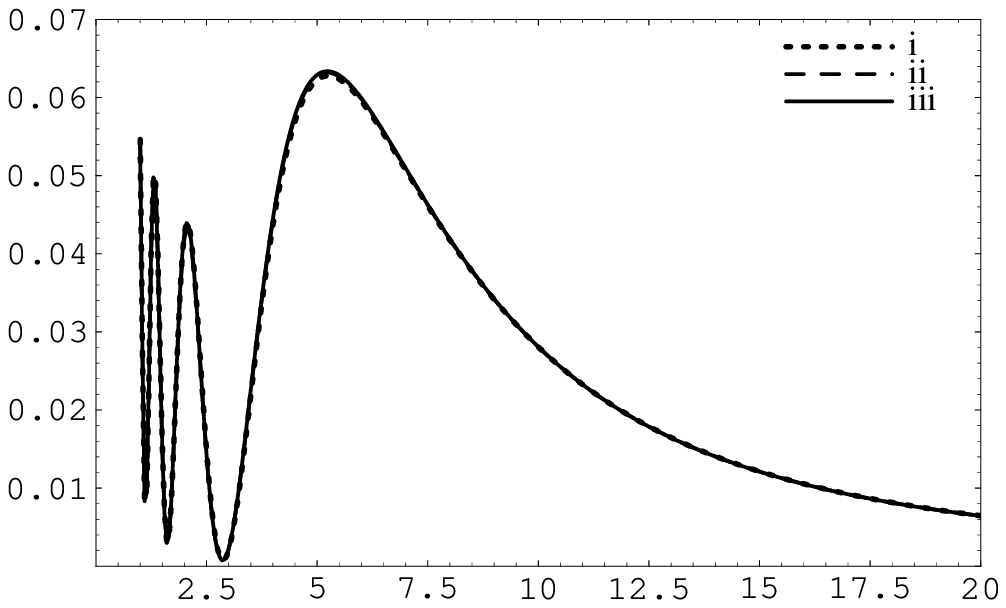}
\end{picture}\\
\vspace{0.4cm}
\caption{Transition probability $P_{\nu_{e} \rightarrow \nu_{\mu}} $ for
$L=3000$km. In this plot we set $\sin\omega=1/\sqrt{2},
\sin\psi=1/\sqrt{2}, \sin\phi=0.1, \Delta m^2_{31}=3\times 10^{-3}{\rm
eV}^2, \Delta m^2_{21}=5\times 10^{-5}{\rm eV}^2$ and
$\delta=\pi/2$. There are three lines corresponding to the matter
profiles; (i) constant density (dotted line), (ii) eq.(\ref{Fourier
Profile}) with N=1 (dashed line), and (iii) PREM matter profile in
Fig.\ref{3000 matter profile} (solid line). These lines are quite
similar to each other so that we cannot see those three lines in the
graph.} 
\label{3000 probability}
\end{figure}

The matter profile\footnote{ To calculate the matter profile, we
linearly interpolate the PREM's data given discretely for the depth. For
this calculation we modify the density of the crust(= most outside
layer) into 2.3 g/cm$^{3} $ from 1 g/cm$^3$.} and its Fourier
coefficients are shown in Fig.\ref{3000 matter profile}. From these
figures we find that the density fluctuation is very small. It should
also be noted that the resonance condition (\ref{Resonance
Condition}) will not be satisfied for any $n$ in neutrino
factories. These facts indicate that the the constant matter density
approximation is valid.  In other words, the Fourier modes will be
irrelevant. To confirm this we plot the oscillation probabilities in
Fig.\ref{3000 probability} with the various matter profiles; (i)
constant density (dotted line), (ii) eq.(\ref{Fourier Profile}) with N=1
(dashed line), and (iii) PREM matter profile in Fig.\ref{3000 matter
profile} (solid line). We find that indeed this figure confirms our
expectation.

\begin{figure}[tbp]
\unitlength 1cm
\begin{picture}(15,9)(0,0)
\put(0,9){\Large Events per GeV}
\put(6,-0.5){(a) $E_\mu=30$ GeV}
\put(14,-0.3){\Large $E_\nu$ [GeV]}
\includegraphics[width=14cm]{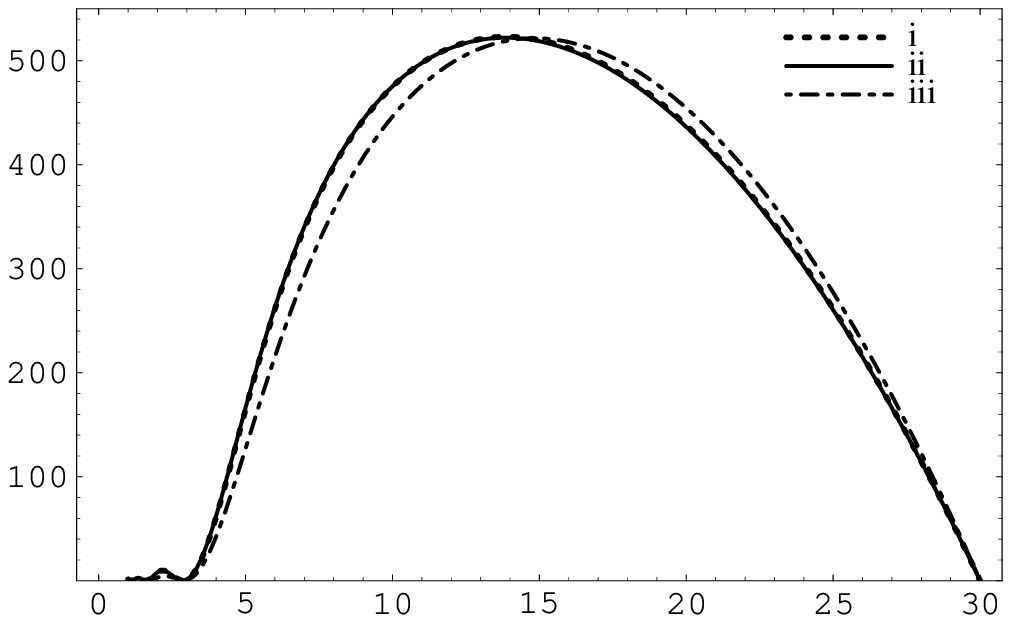}
\end{picture}
\begin{picture}(15,9.6)(0,0)
\put(6,-0.6){(b) $E_\mu=50$ GeV}
\put(14,-0.3){\Large $E_\nu$ [GeV]}
\includegraphics[width=14cm]{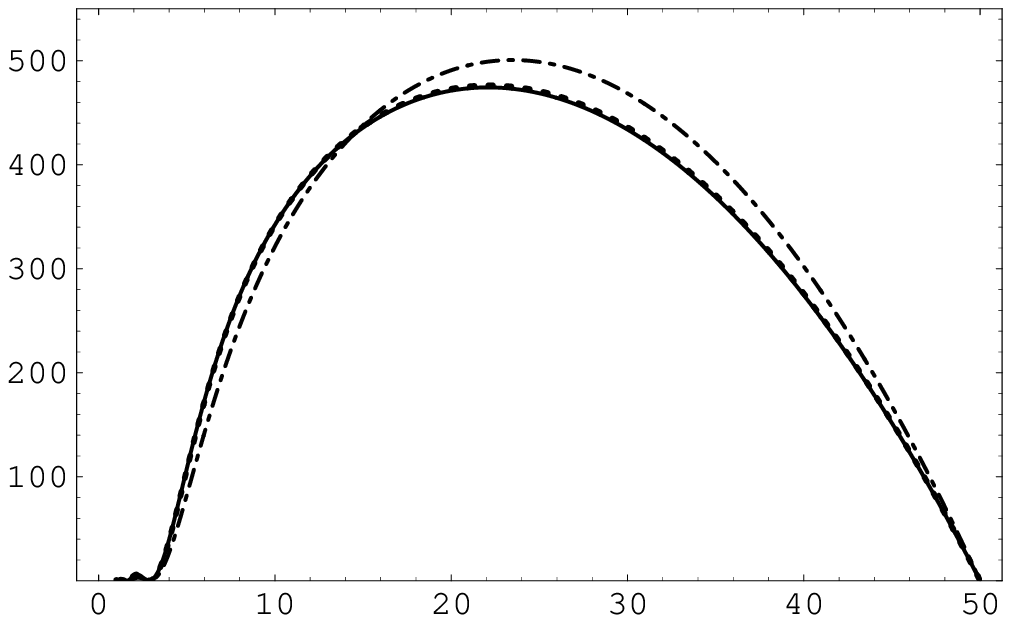}
\end{picture}
\vspace{0.5cm}
\caption{Differential event rate $N_{\nu_e\rightarrow\nu_\mu}$ for
$L=3000$km with (a) $E_\mu=30$ GeV and (b) $E_\mu=50$ GeV. The
oscillation parameters are same as those in Fig.\ref{3000 probability}
except $\delta$. There are three lines with respect to $\delta$ and the
matter profiles;(i) $\delta=\pi/2$ and constant density (dotted line),
(ii) $\delta=\pi/2$ and PREM matter profile (solid line), and (iii)
$\delta=0$ and PREM matter profile (dash-dotted line). The line (iii) is
different from the others.}
\label{3000 event rate}
\end{figure}

The event rates are shown in Fig.\ref{3000 event rate}. We plot the
event rates calculated with the following $\delta$'s and the matter
profiles; (i) $\delta=\pi/2$ and constant density (dotted line), (ii)
$\delta=\pi/2$ and the PREM matter profile (solid line), and (iii)
$\delta=0$ and the PREM matter profile (dash-dotted line). The event
rate with $\delta=\pi/2$ is different from that with $\delta=0$.  We can
observe this difference, if we know all the other theoretical
parameters, the mixing angles, $\Delta m^2$'s and $\bar a$ accurately
enough\footnote{We cannot see this difference, however, since we will
not determine the theoretical parameters accurately
enough\cite{kos}}. Therefore we have to consider the contribution of
$\delta$ to the oscillation probability. However, we cannot see
the difference of the event rate in the case that $a_1=0$ and that
$a_1\ne 0$ with baseline $L=3000$km. Thus we do not have to take into
account the fluctuation of the matter density.

\subsection{ $L=7332$km }
There are many analyses of the oscillation physics with the baseline
length $L=7332$km too. Therefore we consider this case next.

\begin{figure}[tbp]
\unitlength 1cm
\begin{picture}(16,10)(0,0)
\put(0,9.5){\Large$\rho$ [g/cm$^3$]}
\put(6,-0.6){(a) Matter profile}
\put(15.2,0){[km]}
\includegraphics[width=15cm]{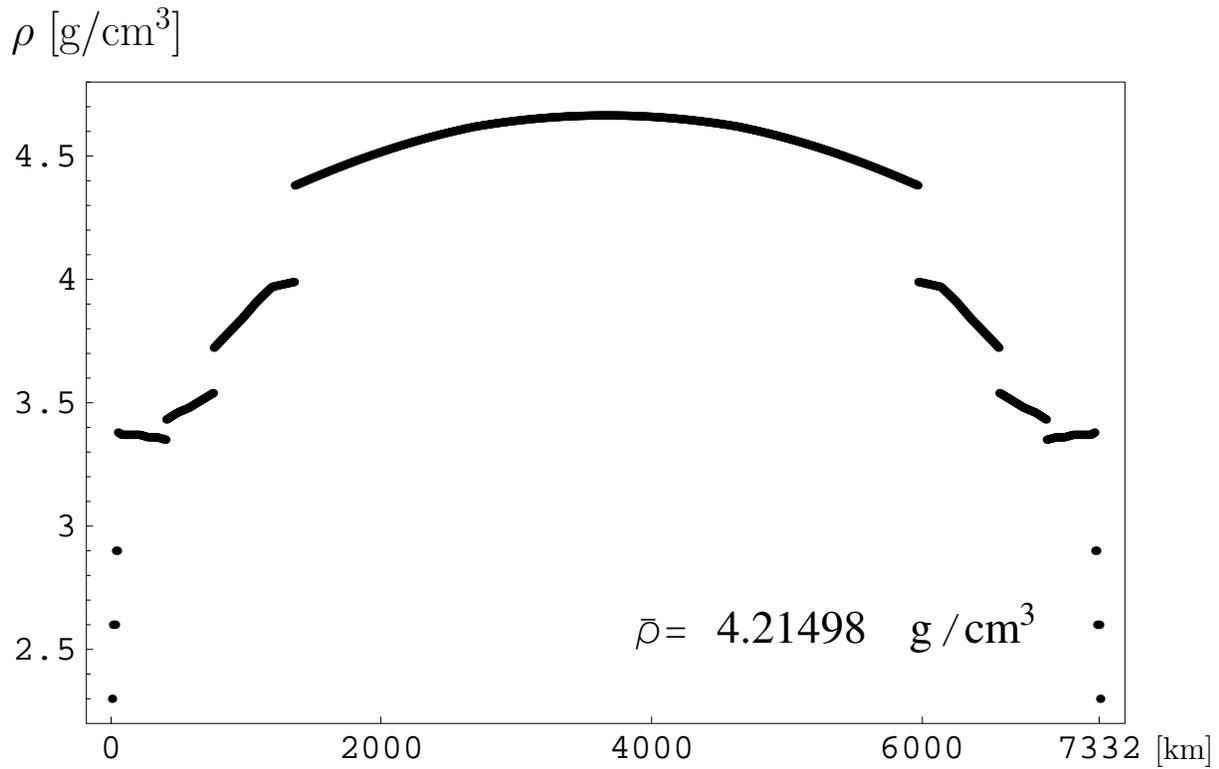}
\end{picture}
\begin{picture}(16,10.5)(0,0)
\put(0,9.5){\Large$\rho_n$ [g/cm$^3$]}
\put(6,-0.6){(b) Fourier Coeeficients}
\put(15.2,0){(mode)}
\includegraphics[width=15cm]{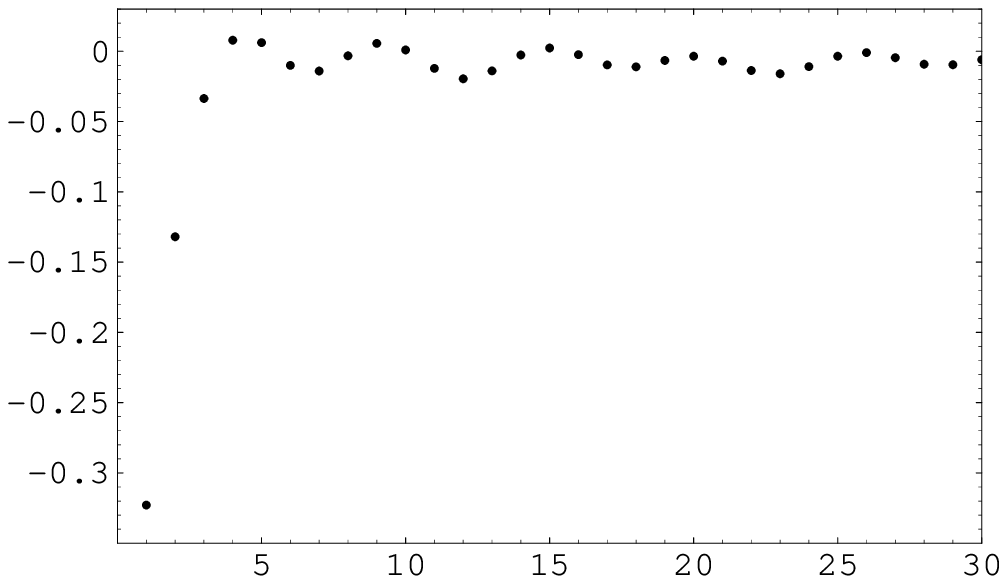} 
\end{picture}
\\
\vspace*{0.5cm}
\caption{(a) Matter profile on the baseline and (b) its Fourier coefficients
 for $L=7332$km.}
\label{7332 matter profile}
\end{figure}

The matter profile based on the PREM and its Fourier coefficients are
shown in Fig.\ref{7332 matter profile}.  We find that the first
two coefficients are large. We argue, however, that the first
Fourier coefficient among them contributes much more to the oscillation
probability because the resonance condition (\ref{Resonance Condition})
for $n=\pm 1$ can be satisfied in a neutrino factory.

\begin{figure}[pht]
\unitlength 1cm
\begin{picture}(16,10)(0,0)
\put(0,9.5){\LARGE$\displaystyle P_{\nu_e\rightarrow\nu_\mu}$}
\put(6.8,-0.8){\LARGE$\displaystyle E_\nu$ [GeV]}
\includegraphics[width=15cm]{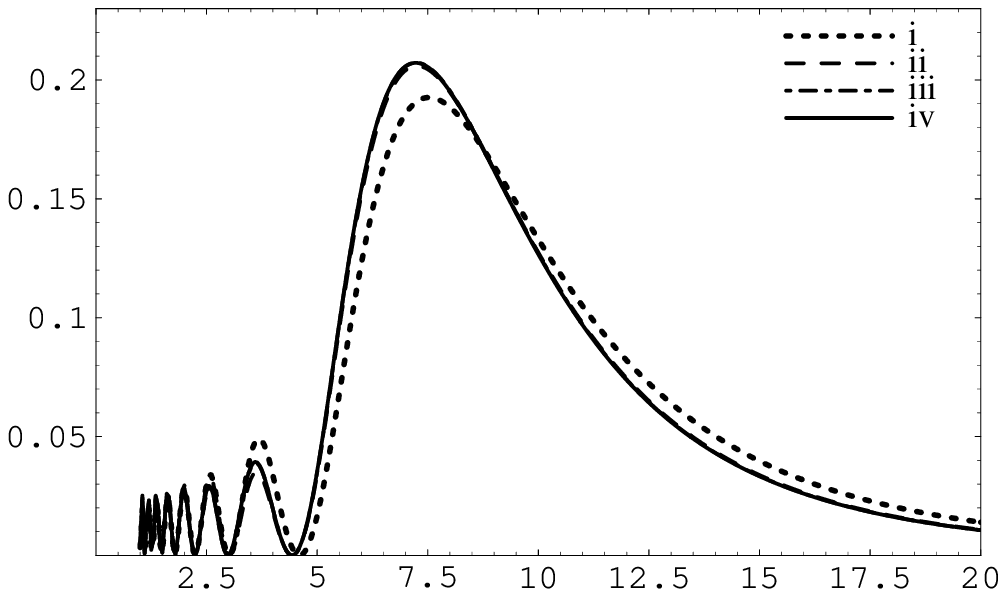} \\
\end{picture}\\
\vspace{0.4cm}
\caption{Transition probability $ P_{\nu_{e} \rightarrow \nu_{\mu}} $
for $L=7332$km.  The oscillation parameters are the same as those in
Fig.\ref{3000 probability}. There are four lines corresponding to the
matter profiles; (i) constant density (dotted line), (ii)
eq.(\ref{Fourier Profile}) with N=1 (dashed line), (iii)
eq.(\ref{Fourier Profile}) with N=2 (dash-dotted line), and (iv) PREM matter
profile in Fig.\ref{7332 matter profile} (solid line). The line (i) is
quite different form the others while the other three lines show quite
similar line shape.}
\label{7332 probability}
\end{figure}

In Fig.\ref{7332 probability}, we compare the oscillation probabilities
calculated with different matter profiles; (i) constant density (dotted
line), (ii) eq.(\ref{Fourier Profile}) with $N=1$ (dashed line),
(iii) eq.(\ref{Fourier Profile}) with $N=2$ (dash-dotted line), and (iv)
the PREM matter profile in Fig.\ref{7332 matter profile} (solid line).

We see that the probability calculated using PREM (Fig.\ref{7332
probability} - (iv)) and another calculated using eq.(\ref{Fourier
Profile}) with $N=1$ (Fig.\ref{7332 probability} - (ii)) are very
similar. However the probability calculated using the constant density
matter profile (Fig.\ref{7332 probability} - (i)) is quite
different. Furthermore there is no significant difference between the
probabilities calculated using eq.(\ref{Fourier Profile}) with $N=1$ and
$N=2$ (Fig.\ref{7332 probability} - (ii) and (iii)). Thus we have to
take into account a new parameter $a_1$ in the analysis of neutrino
factories with the baseline $L=7332$km. This new parameter, $a_1$,
should be estimated by the experimental results.

\begin{figure}[tbp]
\unitlength 1cm
\begin{picture}(15,9)(0,0)
\put(0,9){\Large Events per GeV}
\put(6,-0.5){(a) $E_\mu=30$ GeV}
\put(14,-0.3){\Large $E_\nu$ [GeV]}
\includegraphics[width=14cm]{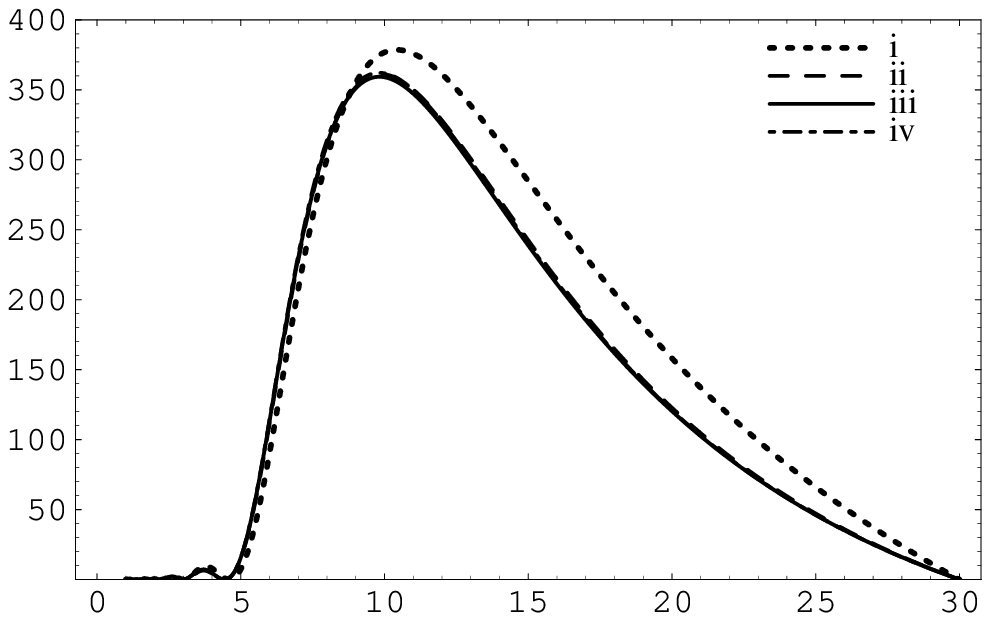} 
\end{picture}
\begin{picture}(15,9.5)(0,0)
\put(6,-0.6){(b) $E_\mu=50$ GeV}
\put(14,-0.3){\Large $E_\nu$ [GeV]}
\includegraphics[width=14cm]{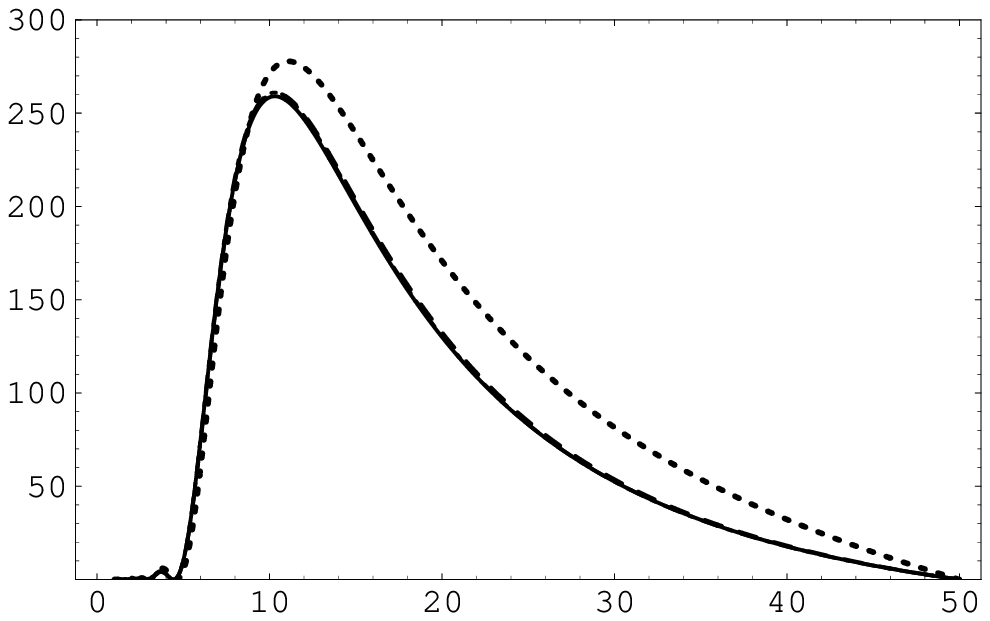}
\end{picture}
\vspace{0.5cm}
\caption{Differential event rate $N_{\nu_e\rightarrow\nu_\mu}$ for
$L=7332$km with (a) $E_\mu=30$ GeV and (b) $E_\mu=50$ GeV. The
oscillation parameters are the same as those in Fig.\ref{3000 probability}
except $\delta$. There are four lines with respect to $\delta$ and the
matter profiles;(i) $\delta=\pi/2$ and constant density ( dotted line),
(ii) $\delta=\pi/2$ and eq.(\ref{Fourier Profile}) with N=1 (dashed
line), (iii) $\delta=\pi/2$ and PREM matter profile (solid line), and
(iv) $\delta=0$ and PREM matter profile (dash-dotted line). The line (i)
is different from the others while the other three are quite similar.}
\label{7332 event rate}
\end{figure}

To see the importance of $a_1$ more clearly, we show the event rate in
Figs.\ref{7332 event rate} using different $\delta$ and different matter
profiles; (i) $ \delta =\pi/2 $ and constant density approximation
(dotted line), (ii) $ \delta =\pi/2 $ and the matter profile
(\ref{Fourier Profile}) with $N=1$ (dashed line), (iii) $ \delta= \pi/2
$ and the PREM matter profile (solid line), and (iv) $ \delta =0 $ and
the PREM matter profile (dash-dotted line).

The distributions of the event rate in (Figs.\ref{7332 event rate}-(i))
and (Figs.\ref{7332 event rate}-(iii)) are quite different from each
other. This means that the estimation of the event rate with constant
density approximation doesn't work.  In other words we cannot estimate
the values of the parameters from the experimental data precisely
without taking into account the contribution of $a_1$. On the other
hand, the event rates in (ii) and (iii) are quite similar. If we can
insist from Figs.\ref{3000 event rate} that we can observe CP violating
effect, then we can insist from Figs.\ref{7332 event rate} that we can
observe the effect of $a_1$. Furthermore since the contribution of $a_1$
cannot be explained by other terms, we expect that we can estimate $a_1$
very well experimentally. We can study geophysics by neutrinos.

Consider the difference between Fig.\ref{7332 event rate}-(iii) and
Fig.\ref{7332 event rate}-(iv). The effect of the CP violating phase on
the event rate is so small that we cannot distinguish between the two
theories that have different CP violating phase.  Indeed the effect of
$\delta$ is smaller than that of $a_2$ which is already beyond the
experimental sensitivity.

\subsection{$L=12000$km}
\begin{figure}[tbp]
\unitlength 1cm
\begin{picture}(16,10)(0,0)
\put(0,9.5){\Large$\rho$ [g/cm$^3$]}
\put(6,-0.6){(a) Matter profile}
\put(15.2,0){[km]}
\includegraphics[width=15cm]{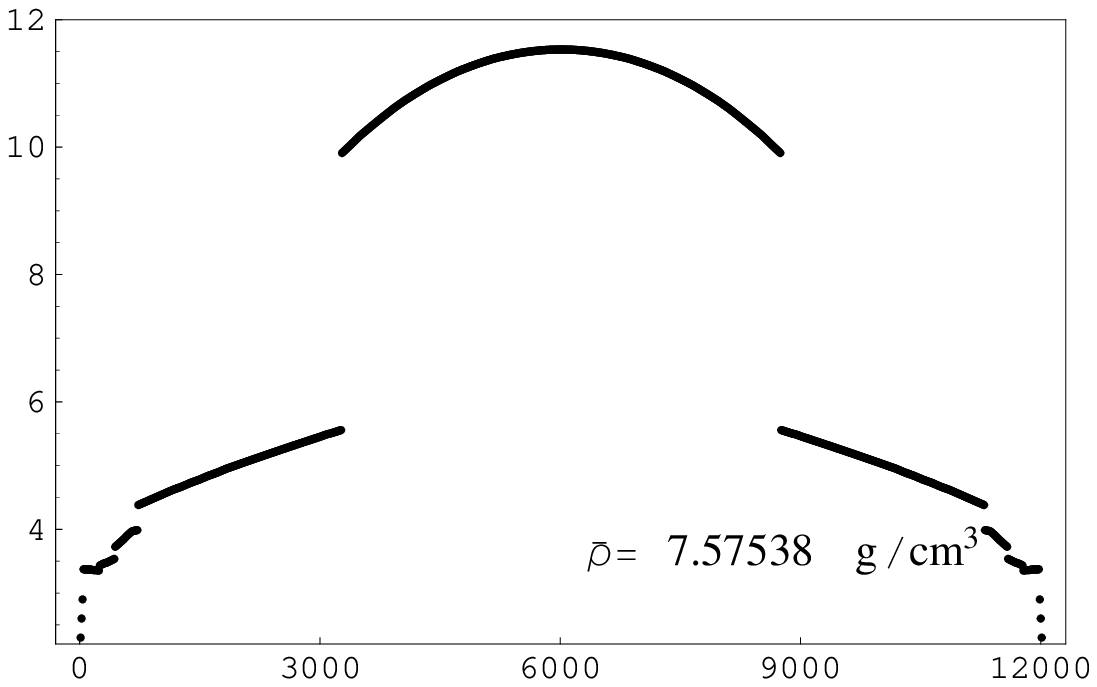}
\end{picture}
\begin{picture}(16,10.5)(0,0)
\put(0,9.5){\Large$\rho_n$ [g/cm$^3$]}
\put(6,-0.6){(b) Fourier Coeeficients}
\put(15.2,0){(mode)}
\includegraphics[width=15cm]{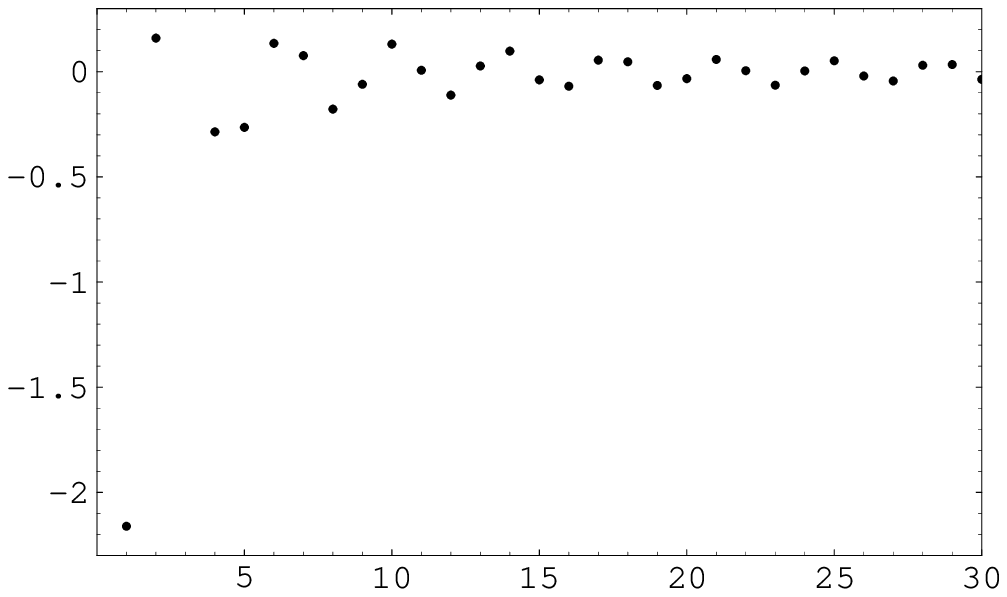} 
\end{picture}
\\
\vspace*{0.5cm}
\caption{(a) Matter profile on the baseline and (b) its Fourier coefficients
 for $L=12000$km.}
\label{12000 matter profile}
\end{figure}

In this case neutrinos penetrate the earth almost along its
diameter. Neutrinos go through both the mantle and the core. Therefore
the matter profile is very complicated. We expect that we can observe
the higher modes in eq.(\ref{Fourier Profile}).

The matter profile and the Fourier coefficients are given in
Fig.\ref{12000 matter profile}. From these figures we expect that not only
$ a_{1} $ but also the higher Fourier modes are relevant.

\begin{figure}[pht]
\unitlength 1cm
\begin{picture}(16,10)(0,0)
\put(0,9.5){\LARGE$\displaystyle P_{\nu_e\rightarrow\nu_\mu}$}
\put(6.8,-0.8){\LARGE$\displaystyle E_\nu$ [GeV]}
\includegraphics[width=15cm]{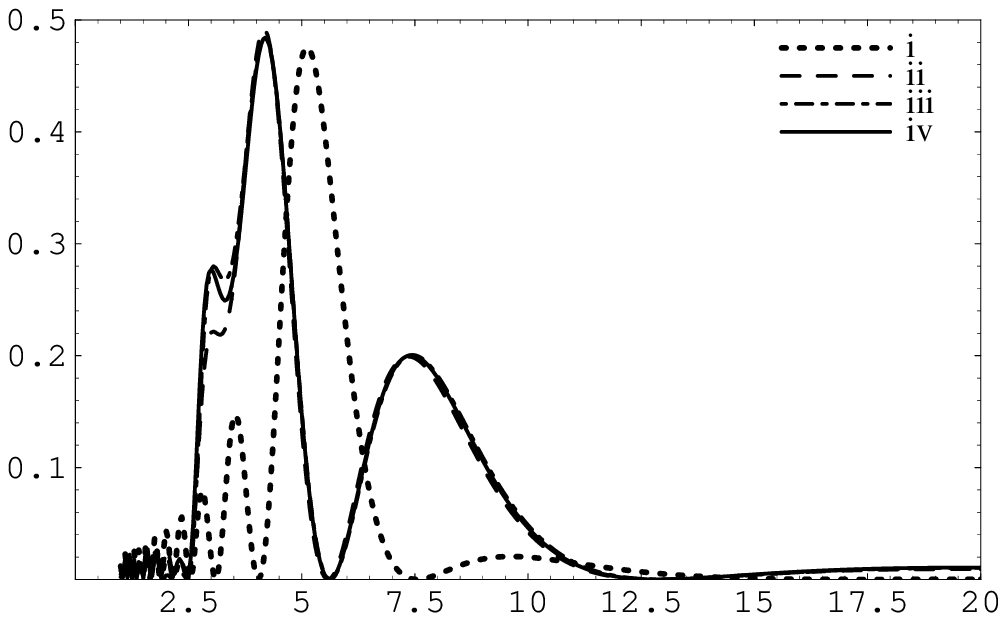} \\
\end{picture}\\
\vspace{0.4cm}
\caption{Transition probability $P_{\nu_{e} \rightarrow \nu_{\mu}} $ for
$L=12000$km. In this plot we use the same oscillation parameters as
those in Fig.\ref{3000 probability}. There are four lines corresponding
to the matter profiles; (i) constant density (dotted line), (ii)
eq.(\ref{Fourier Profile}) with N=1 (dashed line), (iii)
eq.(\ref{Fourier Profile}) with N=3 (dash-dotted line), and (iv) PREM matter
profile in Fig.\ref{12000 matter profile} (solid line). The line (i) is
quite different from the others. The line (ii) is also different from
the line (iV), while the line (iii) shows a quite similar shape with the
line (iv).}
\label{12000 probability}
\end{figure}

\begin{figure}[tbp]
\unitlength 1cm
\begin{picture}(15,9)(0,0)
\put(0,9){\Large Events per GeV}
\put(6,-0.5){(a) $E_\mu=30$ GeV}
\put(14,-0.3){\Large $E_\nu$ [GeV]}
\includegraphics[width=14cm]{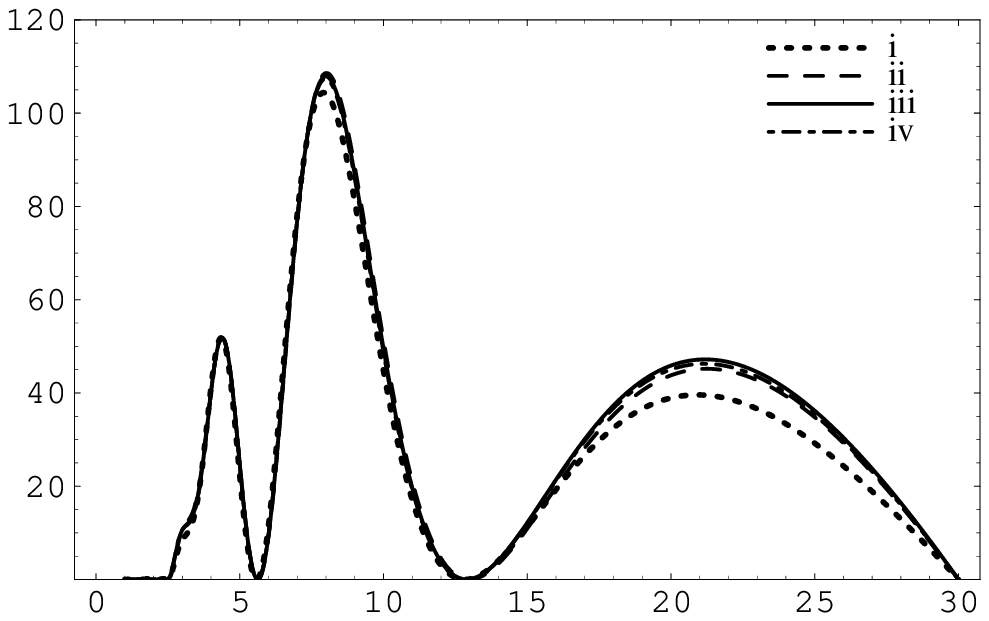}
\end{picture}
\begin{picture}(15,9.5)(0,0)
\put(6,-0.6){(b) $E_\mu=50$ GeV}
\put(14,-0.3){\Large $E_\nu$ [GeV]}
\includegraphics[width=14cm]{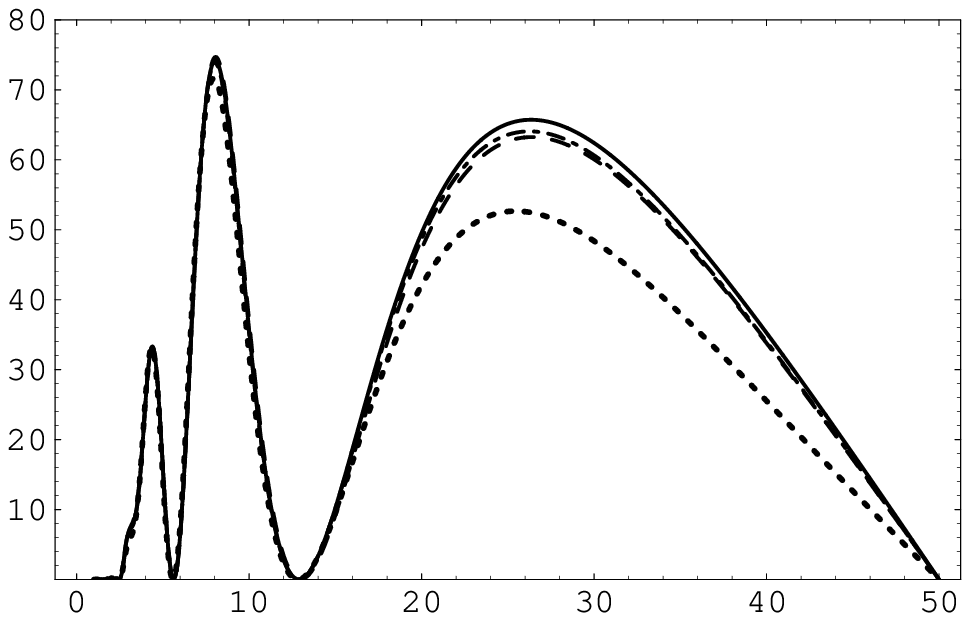}
\end{picture}
\vspace{0.5cm}
\caption{Differential event rate $N_{\nu_e\rightarrow\nu_\mu}$ for
$L=12000$km with (a) $E_\mu=30$ GeV and (b) $E_\mu=50$ GeV. The
oscillation parameters are the same as those in Fig.\ref{3000
probability} except $\delta$. There are three lines corresponding to
$\delta$'s and the matter profiles;(i) $\delta=\pi/2$ and
eq.(\ref{Fourier Profile}) with $N=1$ (dotted line), (ii) $\delta=\pi/2$
and eq.(\ref{Fourier Profile}) with $N=3$ (dashed line), (iii)
$\delta=\pi/2$ and PREM matter profile (solid line); (iv) $\delta=0$ and
PREM matter profile (dash-dotted line).}
\label{12000 event rate}
\end{figure}

In Fig.\ref{12000 probability}, we plot the transition probabilities
corresponding to the matter profiles; (i) constant density (dotted
line), (ii) eq.(\ref{Fourier Profile}) with N=1 (dashed line), (iii)
eq.(\ref{Fourier Profile}) with N=3 (dash-dotted line), and (iv) PREM matter
profile in Fig.\ref{12000 matter profile} (solid line).  From
Fig.\ref{12000 probability} we find that there are significant
contributions to the oscillation probability from the higher modes.  The
probability calculated with the constant density matter profile
(Fig.\ref{12000 probability}-(i)) is quite different from that
calculated with the PREM matter profile (Fig.\ref{12000
probability}-(iv)). Even the probability calculated using the matter
profile eq.(\ref{Fourier Profile}) with N=1 (Fig.\ref{12000
probability}-(ii)) differs apparently from that with the PREM matter
profile (Fig.\ref{12000 probability}-(iv)). On the other hand we find
that the matter profile eq.(\ref{Fourier Profile}) with N=3 mimic very
well the PREM matter profile to calculate the oscillation probability.

We also plot the event rates in Figs.\ref{12000 event rate}.  There the
lines correspond to the following $\delta$ and matter profiles;(i)
$\delta=\pi/2$ eq.(\ref{Fourier Profile}) with $N=1$ (dotted line), (ii)
$\delta=\pi/2$ and eq.(\ref{Fourier Profile}) with $N=3$ (dashed line),
(iii) $\delta=\pi/2$ and PREM matter profile (solid line); (iv)
$\delta=0$ and PREM matter profile (dash-dotted line). We see little
discrepancy in Figs.\ref{12000 event rate}-(ii), (iii), and (iv), while
there is a conceivable difference between Figs.\ref{12000 event
rate}-(i) and the others. We find, therefore that the effect of $\delta$
is irrelevant with the oscillation physics while the first three modes
of the Fourier coefficients have measurable contribution to it.

We examine more carefully whether the higher mode contribution is really
measurable.  The number of the necessary new parameters is quite
dependent on the event rate of the experiment. We can find that there is
conceivable difference between the lines of Figs.\ref{12000 event
rate}-(i) and those of Figs.\ref{12000 event rate}-(ii)\& (iii). However
this difference is not significant statistically in Fig.\ref{12000 event
rate} (a). As long as $N_\mu\times N_{\rm kt}\times \epsilon$ is less
than $5\times 10^{22}$, we need to introduce only $a_1$ as the
theoretical parameter. On the other hand this discrepancy is very
significant in Fig.\ref{12000 event rate} (b). We can measure the
contributions from the higher modes such as $a_2$ and $a_3$. We should
introduce higher modes according to the expected event rate.

\section{Summary and discussion}

We considered how to deal with the matter profile effect.  We proposed
that the Fourier coefficients of the matter profile are used as the
theoretical parameters. Using this method we can evaluate the size of
the ambiguities in the estimate of the mixing parameters.

The perturbative solution for the evolution equation, eq.(\ref{1st order
probability}), implies that the higher Fourier modes are irrelevant.
The introduction of the first few modes gives enough precision to the
estimate of the event rate.

We saw the following three cases in detail numerically.
\begin{itemize}
\item $ L=3000 $km. The matter profile effect itself is
irrelevant. We can assume the matter density is constant.

\item $ L=7332 $km. We need to introduce $a_{1}$ as the theoretical
parameter which should be measured experimentally. On the other hand
the CP violating phase $\delta$ is irrelevant. We cannot determine
the CP violating phase at this baseline.

\item $ L=12000 $km. We need to introduce higher modes as the
theoretical parameters. However, since the event rate with this baseline
length is significantly small, by using only $a_1$ we can estimate
the theoretical parameters accurately within the precision of the
experiment.  Depending on the precision required, we must decide the
number of coefficients to introduce.
\end{itemize}

In general the Fourier modes are relevant ,when the baseline passes
through the lower mantle($L \gtrsim 5000$km).  In this case we need to
use the method developed in this paper to make an accurate analysis. In
other words we can explore the interior of the earth in neutrino
factories.

\subsection*{Acknowledgments}

The authors are grateful to K. Harada and M. Koike for useful discussions.
The work of J.S is supported  in part by a Grant-in-Aid for Scientific
Research of the Ministry of Education, Science and Culture,
\#12047221, \#12740157.


\end{document}